\documentclass[fleqn,usenatbib]{mnras}

\usepackage{newtxtext,newtxmath}

\usepackage[T1]{fontenc}

\DeclareRobustCommand{\VAN}[3]{#2}
\let\VANthebibliography\thebibliography
\def\thebibliography{\DeclareRobustCommand{\VAN}[3]{##3}\VANthebibliography}

\usepackage{graphicx}	
\usepackage{amsmath}	
\usepackage{alphabeta} 

\title[Vela Jr.'s CCO with MUSE]{MUSE observations of the optical nebula surrounding the central compact object in the Vela Junior Supernova Remnant}

\author[J. Suherli et al.]{
Janette Suherli,$^{1}$\thanks{E-mail: suherlij@myumanitoba.ca}
Samar Safi-Harb,$^{1}$
Ivo R. Seitenzahl,$^{2}$
Parviz Ghavamian,$^{3}$
Wynn C. G. Ho,$^{4}$
\newauthor
Chuan-Jui Li,$^{5}$
Ashley J. Ruiter,$^{2}$
Ralph S. Sutherland,$^{6}$
and Fr\'ed\'eric P. A. Vogt$^{7}$
\\
$^{1}$Department of Physics and Astronomy, University of Manitoba, Winnipeg, Manitoba, R3T 2N2, Canada\\
$^{2}$School of Science, University of New South Wales, Australian Defence Force Academy, Canberra, ACT 2600, Australia\\
$^{3}$Department of Physics, Astronomy and Geosciences, Towson University, Towson, MD 21252, USA\\
$^{4}$Department of Physics and Astronomy, Haverford College, 370 Lancaster Avenue, Haverford, PA 19041, USA\\
$^{5}$Institute of Astronomy and Astrophysics, Academia Sinica, No.\ 1, Sec.\ 4, Roosevelt Rd. Taipei 10617, Taiwan\\
$^{6}$Research School of Astronomy and Astrophysics, Australian National University, Cotter Road, Weston Creek, ACT 2611, Australia\\
$^{7}$Federal Office of Meteorology and Climatology – MeteoSwiss, Chemin de l’A\'erologie 1, 1530 Payerne, Switzerland
}

\date{Accepted 2023 November 27. Received 2023 November 27; in original form 2023 October 24}

\pubyear{2023}

\begin{document}
\label{firstpage}
\pagerange{\pageref{firstpage}--\pageref{lastpage}}
\maketitle

\begin{abstract}
Central Compact Objects (CCOs), neutron stars found near the centre of some Supernova Remnants (SNRs), have been almost exclusively studied in X-rays and are thought to lack the wind nebulae typically seen around young, rotation-powered pulsars. We present the first, spatially-resolved, morphological and spectroscopic study of the optical nebula observed at the location of CXOU J085201.4--461753, the CCO in the heart of the Vela Junior SNR. It is currently the only Galactic CCO with a spatially coincident nebula detected at optical wavelengths, whose exact nature remains uncertain. New MUSE integral field spectroscopy data confirm that the nebula, shaped like a smooth blob extending 8$\arcsec$ in diameter, is dominated by [\ion{N}{ii}]$\textnormal{\lambda}\textnormal{\lambda}$6548,6583 emission. The data reveals a distinct and previously unobserved morphology of the H$\textnormal{\alpha}$ emission, exhibiting an arc-like shape reminiscent of a bow shock nebula. We observe a significantly strong [\ion{N}{ii}] emission relative to H$\textnormal{\alpha}$, with the [\ion{N}{ii}]$\textnormal{\lambda}\textnormal{\lambda}$6548,6583 up to 34 times the intensity of the H$\textnormal{\alpha}$ emission within the optical nebula environment. Notably, the [\ion{N}{ii}] and H$\textnormal{\alpha}$ structures are not spatially coincident, with the [\ion{N}{ii}] nebula concentrated to the south of the CCO and delimited by the H$\textnormal{\alpha}$ arc-like structure. We detect additional emission in [\ion{N}{i}], \ion{He}{i}, [\ion{S}{ii}], [\ion{Ar}{iii}], [\ion{Fe}{ii}], and [\ion{S}{iii}]. We discuss our findings in the light of a photoionization or Wolf-Rayet nebula, pointing to a very massive progenitor and further suggesting that very massive stars do not necessarily make black holes.

\end{abstract}




\section{Introduction}

The nature and environment of Central Compact Objects (CCOs) constitute a crucial and fundamental observational test of our understanding of the evolution of massive stars and the theoretical modeling of the supernova (SN) event process in core-collapse explosions. Typified by the CCO first discovered with the Chandra X-ray Observatory in the Cas~A supernova remnant (SNR), CCOs have been identified near the centres of young core-collapse SNRs with no pulsar wind nebulae (PWNe) associations, and have been studied almost exclusively in the X-ray band. Their X-ray spectra are dominated by thermal emission, with blackbody temperatures of 0.2--0.5 keV and typical luminosities of $L_X \sim 10^{33} - 10^{34}$ erg s$^{-1}$ 
(\citealp{Pavlov2004}; \citealp[see][for reviews]{DeLuca2017,SafiHarb2017}). 

Among the 15 known CCOs or CCO candidates\footnote{To access the complete list of Galactic SNRs that host CCOs, users can apply a global filter by typing "CCO" on the SNRcat page: \url{http://snrcat.physics.umanitoba.ca/SNRtable.php} \citep{Ferrand2012}.}, three are detected as pulsars with inferred surface dipole magnetic fields ($\sim$ 10$^{10-11}$ G) and characteristic ages that are several orders of magnitude larger than that of their associated SNRs. As such, CCOs have been designated as anti-magnetars \citep{Gotthelf2013} given their relatively low surface dipole magnetic fields compared to the 10$^{14-15}$~G inferred fields of magnetars. Their unique properties have been attributed to the invoked fall-back from the SN burying their magnetic field and the magnetic field evolution in neutron stars \citep{Ho2011, Rogers2016}. One of the key open questions about CCOs is their SN progenitors. A recent systematic X-ray study of the CCO-hosting ejecta-dominated SNRs showed evidence for low-mass progenitors or a relatively low explosion energy \citep{Braun2019,Braun2023}. In contrast, CXOU J085201.4--461753, the CCO that is investigated in this article, has been suggested to be the product of a massive progenitor \citep{Slane2001,Allen2015}. 

CXOU J085201.4--461753 sits near the heart of Vela Junior \citep[Vela Jr. or G266.2--1.2 or RX J0852.0--4622;][]{Slane2001,Mereghetti2001,Pavlov2001}, a Galactic SNR that was discovered 25 years ago through hard X-ray observations \citep[E $>$ 1.3 keV;][]{Aschenbach1998}. It is predominantly characterized by non-thermal X-ray radiation and has been detected in high-energy gamma-rays \citep[E $\gtrsim$ 100 GeV; e.g.][]{Aharonian2005,Aharonian2007,Abdalla2018}. Vela Jr. remains, thus far, one of the iconic sources in our Galaxy for studying particle acceleration in SNRs. Its age and distance were initially estimated to be $\sim$680 year and $\sim$200 pc, respectively, based on the direct detection of $\gamma$-ray emission attributed to the decay of $^{44}$Ti \citep{Iyudin1998}, which remains controversial \citep[see e.g.][]{Renaud2006}. However, an age of 0.5-1 kyr and a distance of 700 $\pm$ 200 pc, derived from the expansion rate of the remnant \citep{Allen2015}, are now favoured in the literature. 

\begin{figure*}
    \centering
    \includegraphics[width=1.0\textwidth]{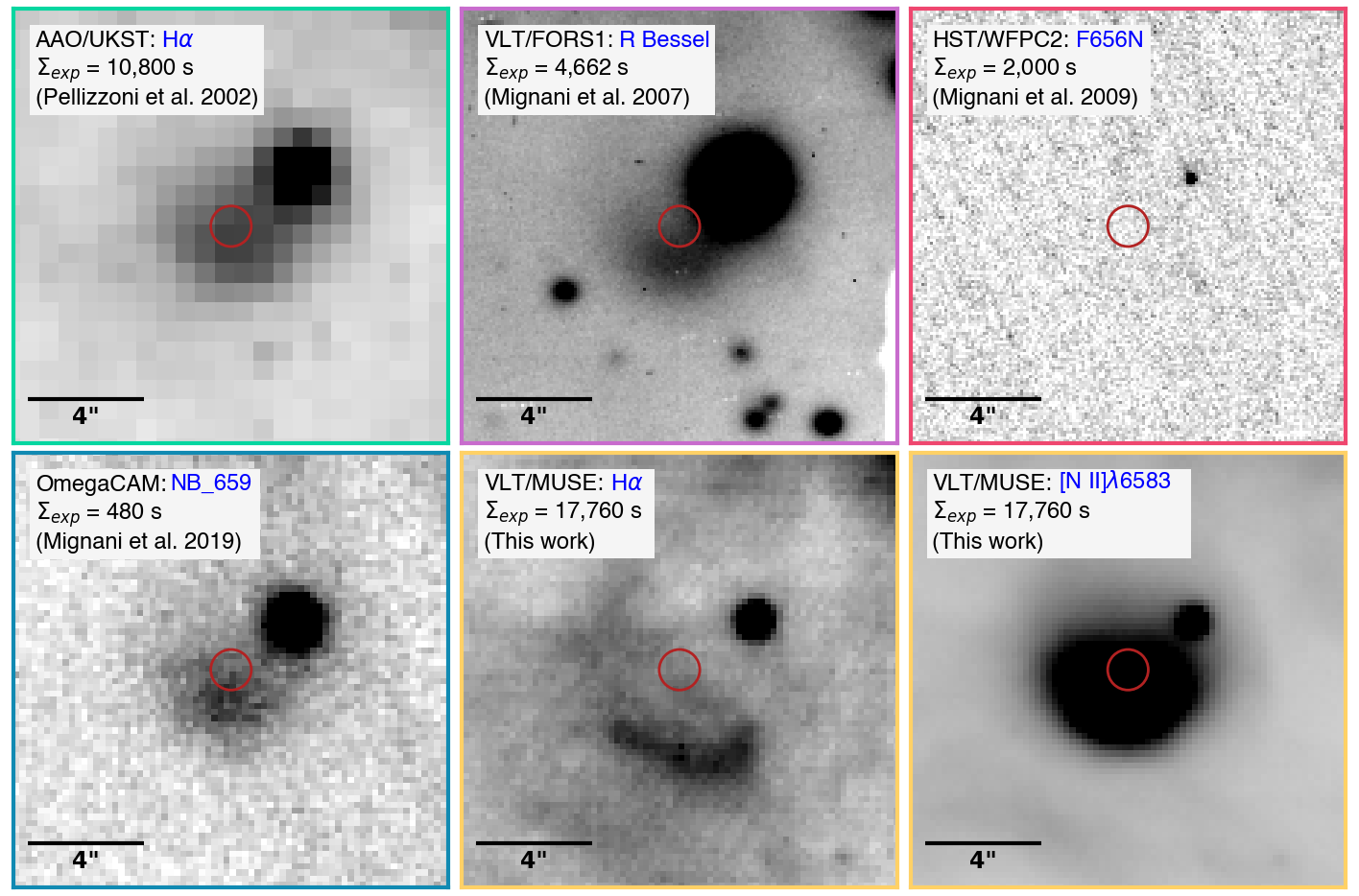}
    \caption{The comparison of 15\arcsec $\times$ 15\arcsec H$\textnormal{\alpha}$ images of the Vela Jr. CCO optical nebula from previous observations and our MUSE observations in H$\textnormal{\alpha}$ and [\ion{N}{ii}]. The red circle marks the CCO location on each panel, and west of the nebula is a field star, Star Z. Each frame colour corresponds to the filter bandpass line colour in Figure \ref{fig:bandpasses}.
    \textbf{Top left:} Digitized image of photographic plate H$\textnormal{\alpha}$ image \citep{Parker1998} from the 3.9 m UK Schmidt Telescope (UKST) at the Australian Astronomical Observation (AAO) showing a nebula with a diameter of roughly 6\arcsec \citep{Pellizzoni2002}.
    \textbf{Top middle:} R-band image from VLT/Focal Reducer Spectrograph (FORS1) showing a kidney-bean shape nebula \citep{Mignani2007}. 
    \textbf{Top right:} HST/WFPC2 F656N observations in which the optical nebula was not detected \citep{Mignani2009}. 
    \textbf{Bottom left:} A serendipitous OmegaCAM \citep[][]{Arnaboldi1998,Kuijken2002,Kuijken2011} observation using the narrow band H\,$\textnormal{\alpha}$ filter NB\_659 \citep{Mignani2019}.
    \textbf{Bottom middle:} H$\textnormal{\alpha}$ integrated image generated from the non-continuum-subtracted MUSE datacube, showing a clear arc-like shape structure.  
    \textbf{Bottom right:} [\ion{N}{ii}] MUSE integrated image, showing a smooth blob-shaped morphology. 
    }
    \label{fig:Ha_images}
\end{figure*}

A particularly enigmatic property of CCOs is their exclusive detection in X-rays, with no confirmed counterparts detected in any other wavelength. This has made the serendipitous discovery of an optical ring of emission in the Small Magellanic Cloud SNR 1E0102.2--7219 surrounding a CCO candidate \citep{Vogt2018} especially remarkable. The true nature of this source however remains uncertain \citep{Hebbar2020,Long2020,Banovetz2021,Li2021}. To date, the Vela Jr. CCO stands as the only Galactic CCO known to have an optical nebular counterpart. The optical nebula was initially identified by \citet{Pellizzoni2002} and was then thought to be H$\textnormal{\alpha}$ emission. It has a roughly circular shape with a diameter of $\sim$6\arcsec (Figure \ref{fig:Ha_images} -- top left), and was proposed to be either a bow-shock nebula produced by the CCO that moves supersonically through the interstellar medium (ISM), or a photo-ionization nebula powered by the radiation from the CCO. Subsequent observations using a broad R-band filter revealed that the shape of the nebula was not exactly spherical, but rather resembles a kidney-bean or an arc-like structure \citep[][Figure \ref{fig:Ha_images} -- top middle]{Mignani2007}. 

Narrow-band imaging observations were conducted using the Wide Field Planetary Camera 2 (WFPC2) on the Hubble Space Telescope (HST) but these failed to detect the optical nebula, down to a 3$\textnormal{\sigma}$ flux limit \citep[][Figure \ref{fig:Ha_images} -- top right]{Mignani2009}. This non-detection was attributed to either the F656N's narrower and bluer bandwidth compared to those of the filters used in previous ground-based observations, or the possibility that the nebula is not dominated by H$\textnormal{\alpha}$ emission after all. Follow-up spectroscopy observations revealed that the spectrum of the nebula is instead dominated by strong and narrow [\ion{N}{ii}]$\textnormal{\lambda}\textnormal{\lambda}$6548,6583, line emissions with only a minor contribution from H$\textnormal{\alpha}$. This led \citet{Mignani2019} to rule out the bow-shock nebula scenario and proposed a plausible association between the nebula and a bright planetary nebula candidate, Wray 16-30 \citep[see][]{Reynoso2006}, located west of the CCO, based on the similarities in their spectra.

In this article, we present the results from our study of the Vela Jr. CCO field with Multi Unit Spectroscopic Explorer \citep[MUSE;][]{Bacon2010}, the integral field spectrograph mounted on the 8.2-m Unit Telescope 4 Very Large Telescope (UT4 VLT) at the Cerro Paranal Observatory in Chile. The MUSE datacubes allowed us to construct spatially-resolved maps of the optical nebula's morphology with sub-arcsecond resolution and derive diagnostic measurements from the spectra. Section \ref{sec:observations} describes our MUSE observations and touches on the data reduction and analysis processes. Our results and characterization of the optical nebula is presented in Section \ref{sec:results}. We discuss our findings in Section \ref{sec:discussions}, and conclude the article in Sect~\ref{sec:summary}.

\section{Observations, Data Reduction, and Post-Processing} \label{sec:observations}

\subsection{MUSE observations}\label{subsec:muse}
We observed the central 2\arcmin\,field of Vela Jr. SNR using MUSE in Wide Field Mode with Ground Layer Adaptive Optics (WFM-AO) configuration \citep{Arsenault2008,Strobele2012} in Service Mode (Program ID 0104.D-0092(B); P.I.: F.P.A. Vogt). MUSE-WFM (nominal) mode covers a spectral range from 4800 \AA\, to 9300 \AA\, and samples the wavelength in 1.25 \AA\,bins at spectral resolutions ranging from R $\sim$ 1770 on the blue side to R $\sim$ 3590 on the red side. It has a field of view of 1 $\times$ 1 arcmin$^{-1}$ with a spatial sampling of 0.2 arcsec pixel$^{-1}$. We arranged our MUSE fields to optimally cover the CCO \citep[RA = 08$^\text{h}$52$^\text{m}$01\fs37, Dec = -46\degr17\arcmin53\farcs50;][]{Mignani2007} optical nebula and most of the structures extending from Wray 16-30 to the northwest, while also avoiding the bright star in the northeast. 

To prevent any background diffuse nebular contamination in the sky frames, we selected an empty sky field, away from the Galactic plane, at 07$^\text{h}$57$^\text{m}$15\fs5; -62\degr41\arcmin00\farcs00. We also applied some position angle changes to each object exposure to minimize the residual background artefacts from the MUSE's 24 identical Integral Field Unit modules. Our observations were carried out in 8 object-sky-pair observing blocks (OBs), for a total of 16 OBs, over 8 nights between 2020 (UT February 25, March 21 and 22\footnote{The program was affected by the COVID-19 closure of the VLT with no science operation on UT4 from March 23, 2020 to November 12, 2020}, December 14, 15, and 17) and 2021 (UT January 10 and 17). Each OB pair consisted of 3 $\times$ 740 s exposure on the object and a 165 s sky exposure. 

\begin{figure*}
    \centering
    \includegraphics[width=1.0\textwidth]{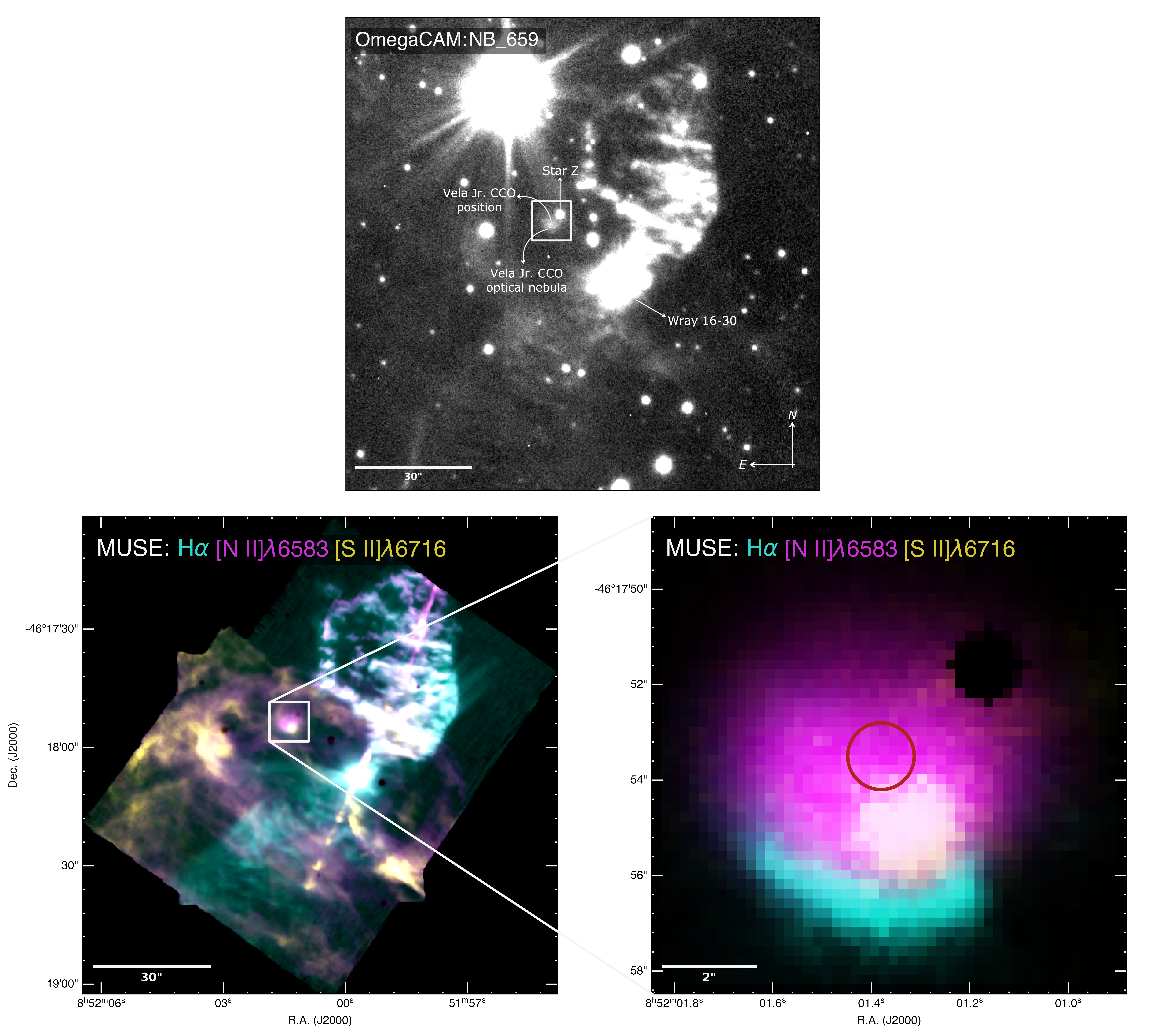}
    \caption{\textbf{Top:} The archival white-light image of the central 2\arcmin\,field of Vela Jr. SNR, obtained with OmegaCAM using the narrow-band H$\textnormal{\alpha}$ filter NB\_659. The Vela Jr. CCO, optical nebula, star Z of \citet{Pavlov2001}, and Wray 16-30 are labeled. The white square indicates the 10\arcsec $\times$ 10\arcsec\,cutout centred on the CCO, corresponding to the area of the bottom right image. 
    \textbf{Bottom left:} The continuum-subtracted image from the MUSE WFM datacube, combining H$\textnormal{\alpha}$ (cyan), [\ion{N}{ii}] (magenta), and [\ion{S}{ii}] (yellow). The image quality of the MUSE datacube is 0.46 arcsec at 5000 Å. Note that the components of the [\ion{N}{ii}] and [\ion{S}{ii}] doublets are resolved spectrally in the MUSE data, and we use only the emission of the brighter components for this figure.
    \textbf{Bottom right:} The continuum-subtracted 10\arcsec $\times$ 10\arcsec\,region, centred on Vela Jr. CCO position (indicated by the red circle), highlighting the spatially resolved optical nebula.
    }
    \label{fig:museRGB}
\end{figure*}

\begin{figure*}
    \centering
    \includegraphics[width=1.0\textwidth]{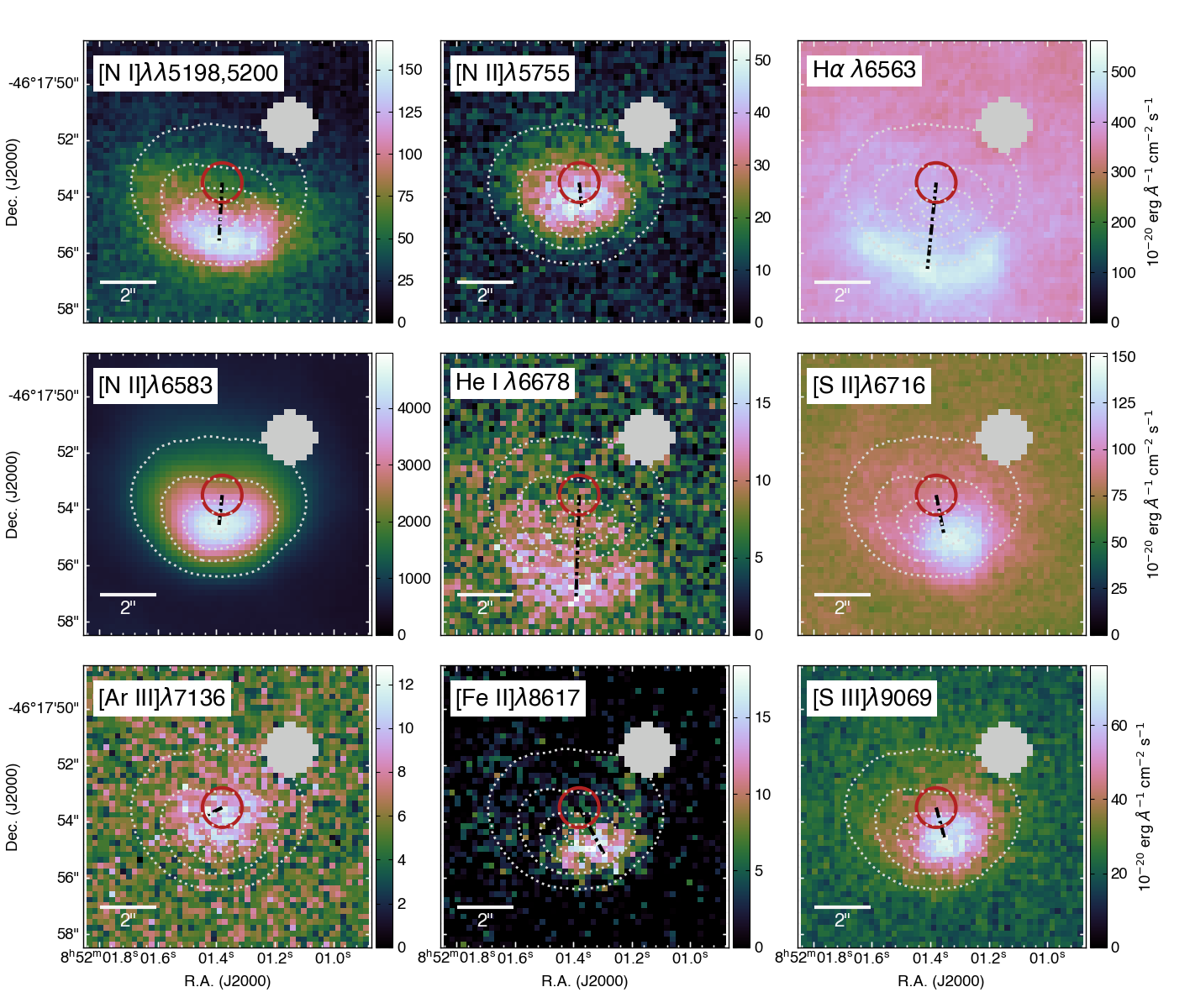}
    \caption{Narrow-band emission line intensity maps of all the bright lines detected in the Vela Jr. CCO optical nebula (from the continuum-subtracted datacube, and the intensity is in units of 10$^{-20}$ erg \AA$^{-1}$ cm$^{-2}$ s$^{-1}$). For [\ion{N}{ii}] and [\ion{S}{ii}] doublets, that are both resolved at the R$\sim$3000 spectral resolution of MUSE, only the maps of the brighter lines are shown. Each map represents the 10\arcsec $\times$ 10\arcsec region centred on the CCO, with north is up, and is overlaid with [\ion{N}{ii}]$\textnormal{\lambda}$6583 intensity contours (white dotted lines). The red circles in all maps mark the positions of the Vela Jr. CCO, and the black dashdot lines indicate the estimated angular distances from the CCO location to the peak of each emission structure (see Table \ref{table:fitResult}). The grey filled-circles in all images are the subtracted region of Star Z, a field star in the area \citep{Pavlov2001}. 
    }
    \label{fig:emissionLineMaps}
\end{figure*}

\subsection{Standard data reduction, calibrations, and post-processing}\label{subsec:datared}
Our MUSE data were reduced individually using the automated MUSE pipeline \citep[v2.8.7;][]{Weilbacher2016, Weilbacher2020} on \textsc{EsoReflex} \citep[ESO Recipe Flexible Execution Workbench v2.11.5;][]{Freudling2013}. We followed the standard reduction procedure for the basic calibration, as well as the sky background subtraction. The output of the MUSE data reduction pipeline is a datacube (with RA, DEC, and wavelength as the three axes) in FITS format. The final datacube has a size of 607 $\times$ 623 spatial pixels (spaxels) = 121.4 $\times$ 124.6 arcsec, which corresponds to a physical size of 0.59 $\times$ 0.60 pc, assuming an upper limit on the distance to the Vela Jr. SNR of 1 kpc \citep{Allen2015}. The image quality of the final datacube is 0.46 arcsec at 5000 Å, sampled with 0.2$\times$0.2 square arcsec spaxels. For the purpose of our analysis in this work, we are focusing on the optical nebula's extent within a 10\arcsec $\times$ 10\arcsec\,area centred on the location of the Vela Jr. CCO.

We corrected the World Coordinate System (WCS) solution of the MUSE datacube using the Gaia Data Release 3 (DR3) catalogue \citep{Gaia2021}. In order to isolate the emission lines from the underlying stellar or nebular continuum in our data and to accurately measure the line properties, we subtracted the continuum from our spectra. First, we fitted the continuum within each spaxel in the datacube using the Locally Weighted Scatterplot Smoothing algorithm (LOWESS; \citealp{Cleveland1979}; the application of LOWESS is described in \citealp{Vogt2017}), and then performed the spaxel-by-spaxel continuum subtractions. We utilized \textsc{brutifus 2019.08.3}\footnote{\url{https://brutifus.github.io/index.html}} \citep{brutifus}, a \textsc{Python} module designed to do post-processing on datacubes from integral fields spectrographs, to execute the WCS correction and continuum subtraction.

\section{Results and Characterization} \label{sec:results}

The top panel of Figure \ref{fig:museRGB} displays the white-light image of the central 2\arcmin\,region of Vela Jr. SNR. This archival image was obtained with OmegaCAM at the VLT Survey Telescope \citep[VST][]{Arnaboldi1998,Kuijken2002,Kuijken2011} in 2015 as part of the VST Photometric H-Alpha Survey of the Southern Galactic Plane and Bulge \citep[VPHAS+][]{Drew2014}, using the narrow-band H$\textnormal{\alpha}$ filter NB\_659. The Vela Jr. CCO position and the optical nebula are labeled, and the white square indicates the 10\arcsec $\times$ 10\arcsec\,cutout centred on the CCO, corresponding to the bottom right image of Figure \ref{fig:museRGB}. Star Z is a field star mentioned in \citet{Pavlov2001}, and Wray 16-30 is a diffuse structure initially designated as a Be star but later reclassified as a planetary nebula \citep{Wray1966,Reynoso2006} and is associated with the star Ve 7-27 \citep[this object was several times improperly referenced as Ve 2-27;][]{Gaia2021}.

The full field of our continuum-subtracted MUSE data as well as the 10\arcsec $\times$ 10\arcsec\,cutout centred on Vela Jr. CCO are shown in the bottom row of Figure \ref{fig:museRGB}, with colours representing the emission from H$\textnormal{\alpha}$ (cyan), [\ion{N}{ii}]$\textnormal{\lambda}$6583 (magenta), and [\ion{S}{ii}]$\textnormal{\lambda}$6716 (yellow). The red circle on the bottom right image indicates the location of the CCO, with a 0.7\arcsec\,radius corresponding to the recomputed Chandra position uncertainty \citep[][]{Mignani2007}. As can be seen from the 10\arcsec\,cutout, the optical nebula is unambiguously dominated by [\ion{N}{ii}] emission and is spatially coincident with the Vela Jr. CCO position. There is a subtle asymmetry toward the south, where the brightness appears slightly enhanced. Additionally, H$\textnormal{\alpha}$ emission is also evident in a distinct arc-like shape, located approximately 3\arcsec\,south of the CCO. The [\ion{N}{ii}] and H$\textnormal{\alpha}$ structures do not spatially align, with the [\ion{N}{ii}] nebula is delimited by the H$\textnormal{\alpha}$ arc-like structure. The [\ion{S}{ii}] emission displays a notably different morphology compared to both the [\ion{N}{ii}] and H$\textnormal{\alpha}$ emissions, appearing as a circular blob structure with a slight off-centre concentration relative to the [\ion{N}{ii}] emission.

\subsection{Nebular morphology in the emission lines} \label{subsec:morphology}
Our MUSE datacube provides the very first spatially resolved view of the optical nebula coincident with the location of Vela Jr. CCO. It is evident that the nebula is predominantly composed of [\ion{N}{ii}] emission. The [\ion{N}{ii}] nebula has a diameter of approximately 8\arcsec, which corresponds to about 0.04 pc, assuming a distance of 1 kpc to the Vela Jr. SNR \citep{Allen2015}. In addition to the bright emissions from H$\textnormal{\alpha}$ $\textnormal{\lambda}$6563 and [\ion{S}{ii}]$\textnormal{\lambda}\textnormal{\lambda}$6716,6731, we also detected emissions from [\ion{N}{i}]$\textnormal{\lambda}\textnormal{\lambda}$5198,5200, [\ion{N}{ii}]$\textnormal{\lambda}$5755, \ion{He}{i} $\textnormal{\lambda}$6678, $\textnormal{\lambda}$7065\footnote{The \ion{He}{i} $\textnormal{\lambda}$5876 emission line is not observable in our MUSE datacube due to the blocking implemented around NaD lines in AO modes observations. This blocking is necessary to prevent contamination from the sodium light used in the laser guide star system.}, [\ion{Ar}{iii}]$\textnormal{\lambda}$7136, [\ion{Fe}{ii}]$\textnormal{\lambda}$8617, and [\ion{S}{iii}]$\textnormal{\lambda}$9069.

In Figure \ref{fig:emissionLineMaps}, we present the narrow-band emission line intensity maps (in erg \AA$^{-1}$ cm$^{-2}$ s$^{-1}$) of the lines detected in the Vela Jr. CCO optical nebula, extracted from the continuum-subtracted MUSE datacube. Each map represents a 10\arcsec $\times$ 10\arcsec\,area, centred on the CCO and was generated by summing the channel images over which the optical nebula structure is seen. The overlaid contour represents the intensity levels of [\ion{N}{ii}]$\textnormal{\lambda}$6583 emission. The red circle in each map marks the CCO position and the black dashdot line indicates the estimated angular distance from the CCO to the peak of emission structure (see column 5 in Table \ref{table:fitResult}). These maps showcase the two distinct morphologies of the optical nebula: 
\begin{enumerate}
    \item An arc-like shape resembling a bow-shock nebula, observed in H$\textnormal{\alpha}$. 
    \item Smooth blob structures, observed in [\ion{N}{ii}], [\ion{S}{ii}], [\ion{Ar}{iii}], [\ion{Fe}{ii}], and [\ion{S}{iii}] emissions, and kidney-bean shaped blob seen in [\ion{N}{i}] and \ion{He}{i} emissions.
\end{enumerate}

\begin{figure}
    \includegraphics[width=\hsize]{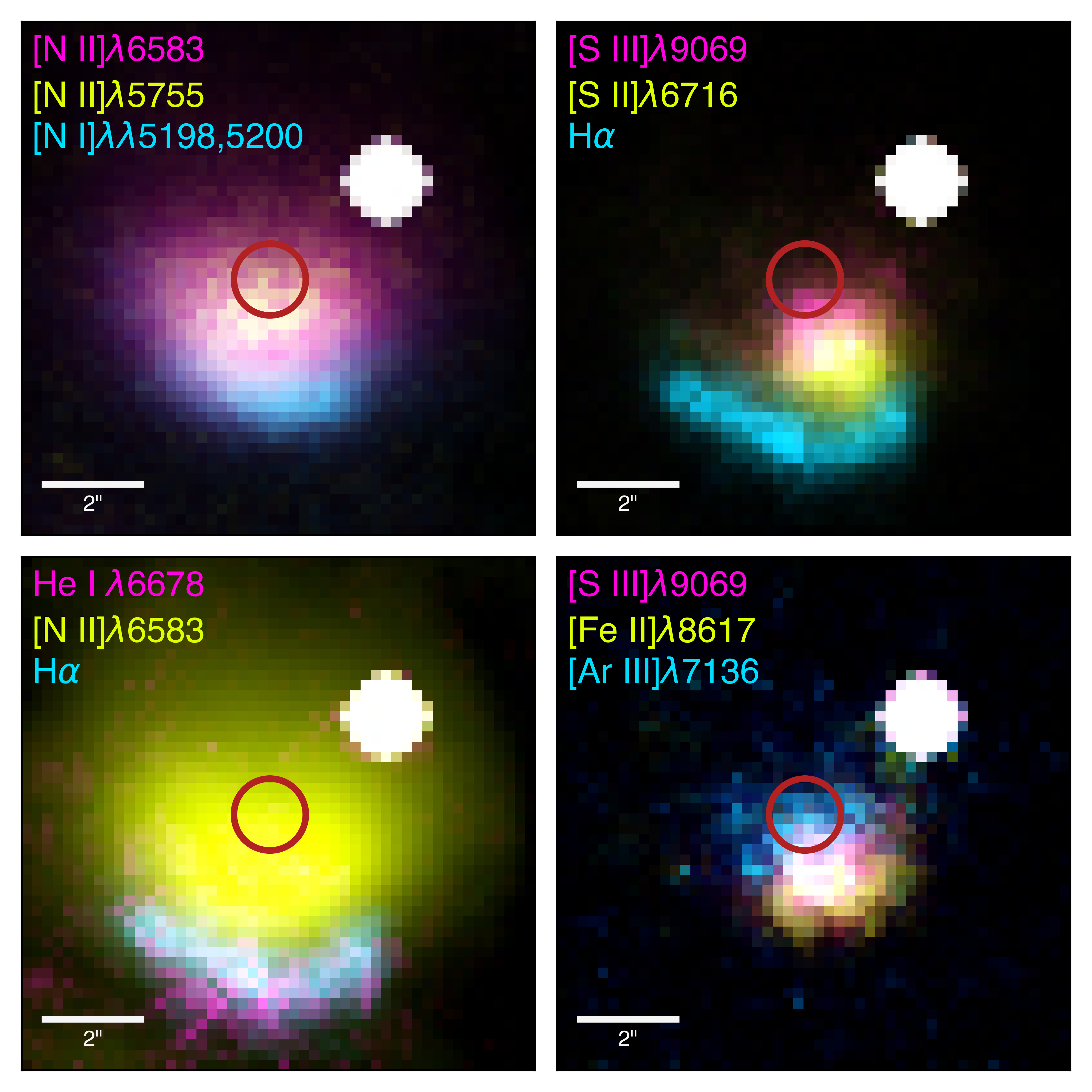}
    \caption{Pseudo-RGB images of different emission lines combinations, with north is up, illustrating the spatial distribution of the emissions detected in the Vela Jr. CCO optical nebula: \textbf{Top left:} [\ion{N}{ii}]$\textnormal{\lambda}$6583, [\ion{N}{ii}]$\textnormal{\lambda}$5755, and [\ion{N}{i}]$\textnormal{\lambda}\textnormal{\lambda}$5198,5200 composite; \textbf{Top right:} [\ion{S}{iii}]$\textnormal{\lambda}$9069, [\ion{S}{ii}]$\textnormal{\lambda}$6716, and H$\textnormal{\alpha}$ composite; \textbf{Bottom left:} \ion{He}{i}$\textnormal{\lambda}$6678, [\ion{N}{ii}]$\textnormal{\lambda}$6583, and H$\textnormal{\alpha}$ composite; \textbf{Bottom right:} [\ion{S}{iii}]$\textnormal{\lambda}$9069, [\ion{Fe}{ii}]$\textnormal{\lambda}$8617, and [\ion{Ar}{iii}]$\textnormal{\lambda}$7136 composite. The red circle in each image marks the position of Vela Jr. CCO. 
    }
    \label{fig:pseudoRGB}
\end{figure}

We created pseudo-RGB images from the continuum-subtracted MUSE datacube to illustrate the spatial distribution of different emission lines detected in the optical nebula, shown in Figure \ref{fig:pseudoRGB}. 
\begin{enumerate}
    \item The first composite, [\ion{N}{ii}]$\textnormal{\lambda}\textnormal{\lambda}$6548,6583, [\ion{N}{ii}]$\textnormal{\lambda}$5755, and [\ion{N}{i}]$\textnormal{\lambda}\textnormal{\lambda}$5198,5200, highlights the different ionization stratification of nitrogen lines and reveals the elongated, kidney-bean-shaped blob of the [\ion{N}{i}] emission tracing the outer edge of [\ion{N}{ii}]. The [\ion{N}{ii}] emissions appear just slightly off-centre from the CCO (Figure \ref{fig:pseudoRGB}--top left). 
    \item The second composite combining [\ion{S}{iii}]$\textnormal{\lambda}$9069, [\ion{S}{ii}]$\textnormal{\lambda}$6716, and H$\textnormal{\alpha}$, shows that both singly- and doubly-ionized sulfur emissions have diffuse circular blob shapes that 
    are near the apex of the H$\textnormal{\alpha}$ emission (Figure \ref{fig:pseudoRGB}--top right). 
    \item The third composite image, which combines \ion{He}{i}$\textnormal{\lambda}$6678 and H$\textnormal{\alpha}$ with respect to [\ion{N}{ii}]$\textnormal{\lambda}$6583. The \ion{He}{i} emission appears to have a kidney-bean shaped blob and traces the south outer edge of [\ion{N}{ii}] emission (Figure \ref{fig:pseudoRGB}--bottom left). 
    \item Lastly, the fourth composite image, combining the emission lines of the products of advanced stages of stellar evolution, [\ion{S}{iii}]$\textnormal{\lambda}$9069, [\ion{Fe}{ii}]$\textnormal{\lambda}$8617, and [\ion{Ar}{iii}]$\textnormal{\lambda}$7136, shows similar circular shapes concentrated near the CCO, offset slightly to the southwest (Figure \ref{fig:pseudoRGB}--bottom right).
\end{enumerate}

We noted that the MUSE datacube reveals a clear absence of structures from the H$\textnormal{\textnormal{\beta}}$ and [\ion{O}{iii}]$\textnormal{\lambda}\textnormal{\lambda}$4959,5007 emissions within the Vela Jr. CCO optical nebula. 

\subsection{Spectral analysis of the optical nebula} \label{subsec:spectra}
\begin{figure*}
    \includegraphics[width=0.9\textwidth]{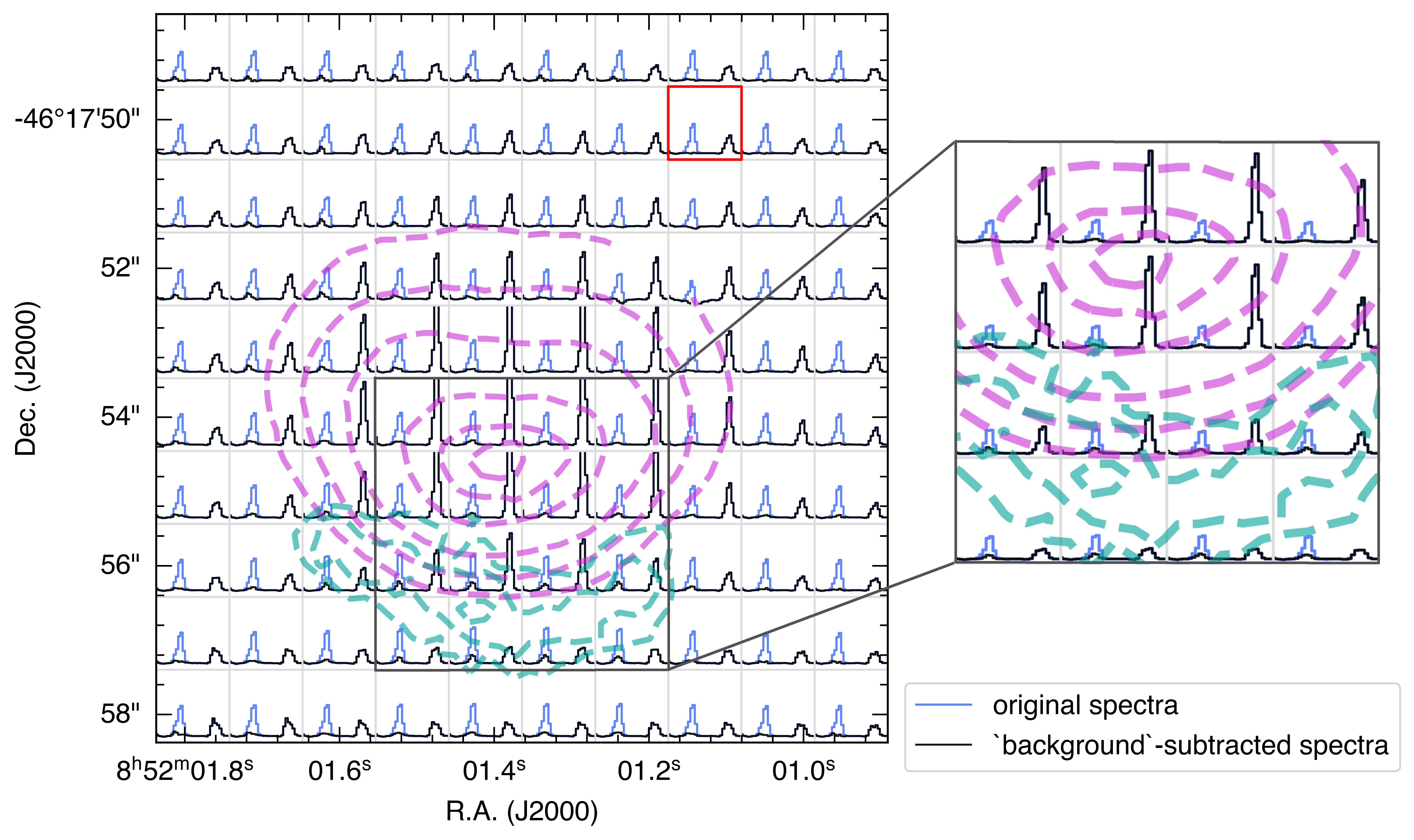}
    \caption{H$\textnormal{\alpha}$ and [\ion{N}{ii}]$\textnormal{\lambda}$6583 line profiles from 6555 \AA\, to 6590 \AA, obtained by integrating over 5 $\times$ 5 spaxels boxes (1\arcsec $\times$ 1\arcsec) and overplotted on H$\textnormal{\alpha}$ (cyan dotted-line) and [\ion{N}{ii}] (magenta dotted-line) intensity contour maps. Within each box, the blue and black lines respectively represent the spectra from MUSE data and the H$\textnormal{\alpha}$ emission background-corrected spectra. The red rectangle indicates the region from which the line profile spectra were used for the correction. On the right, a magnified snippet shows the full extent of the spectral lines strengths. 
    }
    \label{fig:overplot}
\end{figure*}

Figure \ref{fig:overplot} shows the intensity contour maps of the optical nebula morphology in H$\textnormal{\alpha}$ (cyan) and [\ion{N}{ii}] (magenta), superimposed with the spectral line profiles from 6540 \AA\, to 6590 \AA, covering the H$\textnormal{\alpha}$ and [\ion{N}{ii}]$\textnormal{\lambda}$6583 lines (the blue and black lines). These line profiles were generated by integrating the spectra over 5 $\times$ 5 spaxels boxes (1\arcsec $\times$ 1\arcsec). The line profile of H$\textnormal{\alpha}$ appears fairly constant in strength across the spaxel boxes, shown as the blue lines in each box. This suggests the presence of some consistent background H$\textnormal{\alpha}$ contribution across the field. To correct the H$\textnormal{\alpha}$ line profile for our analysis, we selected an area outside the optical nebula where there is no significant emission from the H$\textnormal{\alpha}$ nebula, as indicated by the red rectangle in Figure \ref{fig:overplot}. We then used the average line profile from this region for the  correction. The black lines represent the corrected H$\textnormal{\alpha}$ line profile, and the snippet on the right provides a magnified view of the line profiles, showing the full extent of the [\ion{N}{ii}]$\textnormal{\lambda}$6583 line strength.

We fitted the integrated spectrum of each emission line detected in the optical nebula individually with a single-component Gaussian function, with the exception for the [\ion{N}{ii}]$\textnormal{\lambda}\textnormal{\lambda}$6548,6583 and [\ion{S}{ii}]$\textnormal{\lambda}\textnormal{\lambda}$6716,6731 where each was fitted jointly. We imposed the same velocity and line width for both emission line doublets and constrained the flux ratio [\ion{N}{ii}]$\textnormal{\lambda}$6583/$\textnormal{\lambda}$6548 to be equal to the theoretical value of $\sim$3 \citep{Osterbrock1985}. The line fitting was conducted on the integrated spectra using our own custom \textsc{Python} routine. It starts by determining the spatial region for spectra integration, which is chosen based on the intensity level of each emission line's nebular structure. A non-linear least square optimization method was then employed to fit a single-component Gaussian to the integrated spectrum. The uncertainties of the fitted parameters were measured as one standard deviation errors. The resulting measurements, which include the total observed line flux, radial velocity, and the corrected full width at half maximum (FWHM) velocity dispersion, accounting for the instrument line broadening of MUSE, were derived for all prominent lines and are presented in Table \ref{table:fitResult}. Additionally, we estimated the angular distance from the CCO location to the peak emission location of each line emission through visual inspection (the last column of Table \ref{table:fitResult} and indicated with the black dashdot lines in Figure \ref{fig:emissionLineMaps}).

The derived radial velocities of most of the emission lines indicate that the nebular structures are moving toward us, with the exception for the H$\textnormal{\alpha}$ and [\ion{S}{iii}]. From the integrated spectra, we measured an unusually high [\ion{N}{ii}]6548$+$6583/H$\textnormal{\alpha}$ value of 33.8 for the optical nebula structure. 

Using the measurements from the integrated spectra presented in Table \ref{table:fitResult}, we determined the electron temperature and electron density in the optical nebula utilizing the [\ion{N}{ii}] ($\textnormal{\lambda}$6548+$\textnormal{\lambda}$6583)/$\textnormal{\lambda}$5755 and the [\ion{S}{ii}]$\textnormal{\lambda}$6716/$\textnormal{\lambda}$6731 flux ratios. We derived these values using the PyNeb code \citep{Luridiana2015} which solves the equilibrium equations for an n-level atom and works iteratively with the ratio of the [\ion{N}{ii}] lines, as well as of the [\ion{S}{ii}] lines to obtain more accurate values. The [\ion{N}{ii}] lines provide an estimate for the mean electron temperature of 
7800 K for the optical nebula, and the [\ion{S}{ii}] lines give an electron density value of 35 cm$^{-3}$.

\begin{table*}
\caption{Observed total (integrated) line fluxes, radial velocities, derived full width at half maximum (FWHM) velocities that are corrected for the MUSE's instrument line broadening, and the angular distances estimated visually from the Vela Jr. CCO location to the peak emission of each line emission. The uncertainties of the fitted parameters were measured as one standard deviation error. NOTE -- We did not perform the fitting for [\ion{Fe}{ii}]$\textnormal{\lambda}$8617, and \ion{He}{i} $\textnormal{\lambda}$7065 due to their low integrated flux level.}
\label{table:fitResult}
\begin{tabular}{lcccc}
\hline
Line & \textit{f}$_\textnormal{\lambda}$,obs & v & FWHM (corrected) & Angular Distance \\
 & (10$^{-16}$ erg cm$^{-2}$ s$^{-1}$) & (km s$^{-1}$) & (km s$^{-1}$) & (\arcsec)\\
\hline
\hline
[\ion{N}{i}]$\textnormal{\lambda}$5197 & 2.76 & -2.6 & 44.5 & 2.24 \\\relax
[\ion{N}{i}]$\textnormal{\lambda}$5200 & 3.39 & -2.6 & 44.3 & 2.24 \\\relax
[\ion{N}{ii}]$\textnormal{\lambda}$5755 & 1.50 & -12.1 & 61.9 & 1.02 \\\relax
[\ion{N}{ii}]$\textnormal{\lambda}$6548 & 66.36 & -9.0 & 54.6 & 1.26 \\\relax
H$\textnormal{\alpha}$ $\textnormal{\lambda}$6563 & 8.17 & 3.3 & 81.7 & 3.26 \\\relax
[\ion{N}{ii}]$\textnormal{\lambda}$6583 & 209.84 & -9.0 & 54.3 & 1.26 \\\relax
\ion{He}{i} $\textnormal{\lambda}$6678 & 1.14 & -2.6 & 87.1 & 3.82 \\\relax
[\ion{S}{ii}]$\textnormal{\lambda}$6716 & 13.76 & -23.3 & 163.6 & 1.6 \\\relax
[\ion{S}{ii}]$\textnormal{\lambda}$6731 & 9.70 & -23.3 & 163.3 & 1.6 \\\relax
[\ion{Ar}{iii}]$\textnormal{\lambda}$7136 & 1.51 & -0.3 & 136.7 & 0.89 \\\relax
[\ion{S}{iii}]$\textnormal{\lambda}$9069 & 3.79 & 17.8 & 71.2 & 1.20 \\
\hline
\end{tabular}
\end{table*}

\subsection{A CCO-ionized optical nebula} \label{subsec:cco-nebula} 
Figures \ref{fig:museRGB}--\ref{fig:pseudoRGB} of the inner 4--6 arcseconds optical nebula show unusual emission, including very strong [\ion{N}{ii}] emission relative to H$\textnormal{\alpha}$, as well as diffuse [\ion{S}{ii}] and [\ion{S}{iii}]. The undetected (or weak) H$\textnormal{\textnormal{\beta}}$ is also suggestive of a non-standard H$\textnormal{\alpha}$/H$\textnormal{\textnormal{\beta}}$ ratio, which makes the common Balmer decrement reddening estimation problematic. However, it is often the case in photoionized nebulae modeling, such as fast shock precursors and other partially ionized plasmas \citep{Sutherland2017,Dopita2017} for the H$\textnormal{\alpha}$/H$\textnormal{\textnormal{\beta}}$ ratio to be intrinsically greater than the canonical 2.86 Case B (optically thick nebula where the recombinations of H atom to the ground state generate photons that are absorbed locally) ratio and H$\textnormal{\alpha}$ can be enhanced.

We performed a preliminary photoionization modeling with MAPPINGS~V v5.2.0\footnote{\url{https://mappings.anu.edu.au}} \citep{mappings} using a Wolf-Rayet (WR) star progenitor -- the late phase of massive star evolution identified by high mass-loss rate and fast stellar winds. As a hypothesis, we took a WN star (a subtype of WR stars, characterized by strong helium and nitrogen lines) surface composition of a star with an initial mass of 100 M${\sun}$ from \citet{Roy2020} and evolved it past the late-type WN (WNL) stage to the full He core burning WN of 50--60\,M$_{\sun}$. By this late stage of evolution, the atmosphere and surface composition are insensitive to the initial abundances of the star, dominated by helium and nitrogen, and are heavily CNO-processed. Under the assumption that these compositions from the last 500,000\,years of the stellar lifetime could constitute a late stage WN wind environment for any hot CCO remnant, we constructed a simple spherical isobaric X-ray nebula model using MAPPINGS~V to test the general feasibility.

\begin{figure*}
    \centering
    \includegraphics[width=0.7\textwidth]{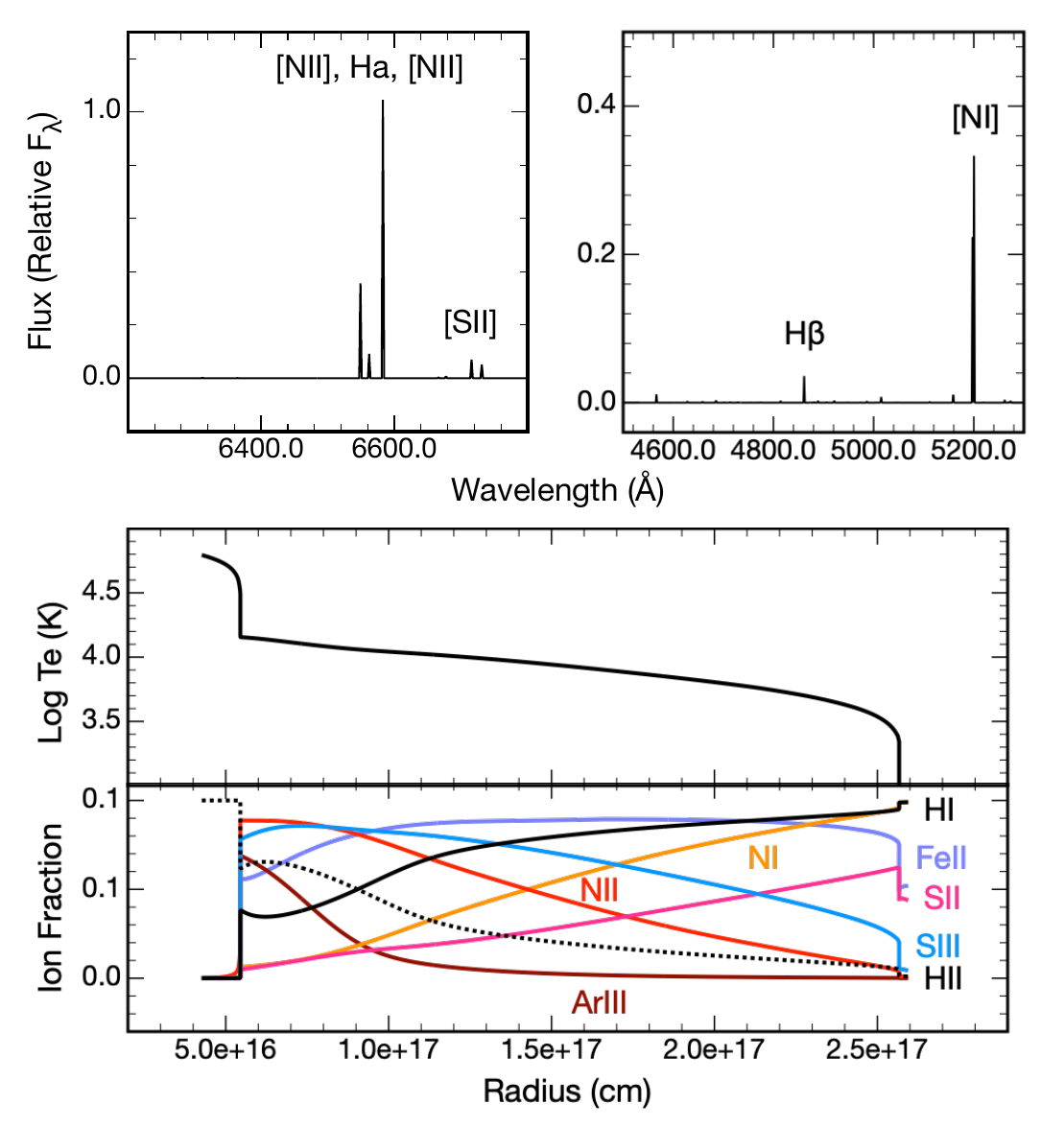}
    \caption{MAPPINGS~V test model of a nebula with 60\,M$_{\sun}$ WN progenitor. \textbf{Top:} The simulated spectra of the nebula. \textbf{Bottom:} The intrinsic variation of electron temperatures for a spherical model of the nebula as a function of the nebular radius and the spherical MAPPINGS~V model of the ionization structures within the nebula.}
    \label{fig:mappingstest}
\end{figure*}

Using the \citet{Slane2001} estimated 0.47 keV blackbody temperature for the Vela Jr. CCO's thermal X-ray emission, a $5.45 \times 10^6$\,K blackbody as a uniform 10\,km radius source was placed in the WN stage with gas composition listed in Table \ref{table:modelm60nebula}. The pressure of the nebula was iterated until it yielded the densities that satisfy the observed [\ion{S}{ii}]$\textnormal{\lambda}$6731/$\textnormal{\lambda}$6716 density ratio. The result of our test model is summarized in Figure \ref{fig:mappingstest}. The resulting nebula has a $\sim 2.5\times10^{17}$\,cm radius in spherical symmetry and displays a high [\ion{N}{ii}]/H$\textnormal{\alpha}$ ratio, an extended [\ion{S}{ii}], and no [\ion{S}{iii}]$\textnormal{\lambda}$6312 line (as the nebula was largely too cool) in its spectra, along with a strong [\ion{N}{i}] line, as shown in the top panels of Figure \ref{fig:mappingstest}. The inner $5\times10^{16}$\,cm of this nebula is very hot and highly ionized, producing no observable optical nebular lines. The nebula rapidly cools in an extended region spanning 0.5--2.5$\times10^{17}$\,cm, with temperatures from 3 $\times 10^3$ to 1.5$\times 10^4$\,K, and emits observable optical lines (Figure \ref{fig:mappingstest} bottom panels).

\begin{table*}
\caption{Gas composition of 60\,M$_{\sun}$ WN star progenitor for MAPPPINGS~V test model}
\label{table:modelm60nebula}
\begin{tabular}{ccc|ccc}
\\
\hline
                            &  {Number}   & {Mass}   & &{Number}  &  {Mass}     \\
{Element} &  {$\log(n_H)=0.0$} & {Fraction} & {Element}&{$\log(n_H)=0.0$} & {Fraction}  \\
\hline
\hline
{H}       & 0.00        & 2.26E-01 & {Al} & -4.90 & 7.59E-05 \\
{He}      & -0.07       & 7.64E-01 & {Si} & -4.12 & 4.81E-04 \\
{C}       & -4.58       & 7.03E-05 & {Si} & -4.50 & 2.30E-04 \\
{N}       & -2.76       & 5.53E-03 & {Cl} & -5.25 & 4.49E-05 \\
{O}       & -5.06       & 3.14E-05 & {Ar} & -4.35 & 4.02E-04 \\
{Ne}      & -3.74       & 8.32E-04 & {Ca} & -5.25 & 5.03E-05 \\
{Na}      & -4.46       & 1.77E-04 & {Fe} & -4.29 & 6.39E-04 \\
{Mg}      & -4.14       & 3.95E-04 & {Ni} & -4.65 & 2.92E-04 \\
\hline
\end{tabular}
\end{table*}

This preliminary result is circumstantial, and more detailed modeling that includes partial CCO surface heat and non-thermal components will be presented in future work. Nevertheless, it supports the existence of a CCO, and suggests that at least some of the nebulosity is attributed to an X-ray blackbody object, in addition to the widespread shock processes. The general ionization structure of the nebula from the test model broadly agrees with the observations, specifically in reproducing the strong widespread of nitrogen and the extended sulfur emission in particular, which distinguished this from a typical H II region. With further exploration and modeling, this hypothesis may resolve if the inner 4\arcsec is a neutron-star-photoionized nebula or bubble, and the bright H$\textnormal{\alpha}$ arc to the south is an interaction or bow shock structure. If it is true that the CCO is a $\sim5.5$\,MK, with about 10\,km radius source, and able to photoionize a WN composition circum--progenitor environment this lends support to the \citet{Allen2015} suggestion that Vela Jr has a massive progenitor.

\section{Discussion} \label{sec:discussions}

\subsection{Comparison of MUSE observations with previous H$\textnormal{\alpha}$ observations} \label{subsec:ha}
\begin{figure}
    \includegraphics[width=\hsize]{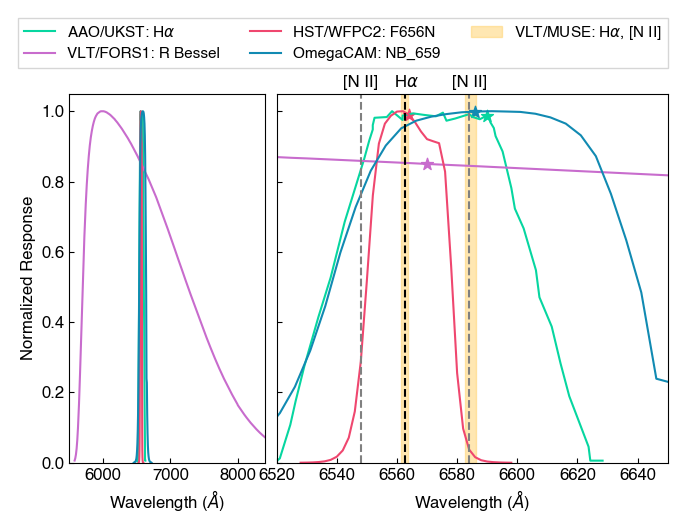}
    \caption{The filter bandpass response range of all the filters used in the previous optical observations of the Vela Jr. CCO optical nebula is represented by coloured lines, that match the frame colour of the images in Figure \ref{fig:Ha_images}. The MUSE wavelength ranges from which each MUSE image is integrated are shown as yellow shaded areas. \textbf{Left:} The filter bandpasses spanning from 5400 \AA\, to 8400 \AA, to show the extended range of the R Bessel filter used for the VLT/FORS1 observations. \textbf{Right:} The filter bandpasses range from 6520 \AA\, to 6650 \AA, to emphasize the narrow-band filters. Star symbols indicate the central wavelength of each filter}
    \label{fig:bandpasses}
\end{figure}

We compiled previously published H$\textnormal{\alpha}$ observations of the optical nebula from \citet{Pellizzoni2002}, \citet{Mignani2007}, \citet{Mignani2009}, and \citet{Mignani2019}, alongside the MUSE images in H$\textnormal{\alpha}$ and [\ion{N}{ii}] from the non-continuum-subtracted datacube in Figure \ref{fig:Ha_images}. Each frame colour corresponds to the line colour of the respective filter bandpass range in Figure~\ref{fig:bandpasses}. Since its discovery in 2002, the optical nebula spatially coincident at the position of the CCO in Vela Jr. SNR was initially believed to be H$\textnormal{\alpha}$ emission, with a kidney-bean-like shape. MUSE data unequivocally reveal that the observed shape of the optical nebula is primarily comprised of [\ion{N}{ii}] emission rather than H$\textnormal{\alpha}$, and the H$\textnormal{\alpha}$ emission exhibits a distinct arc-like morphology. This could be attributed to the fact that we observe a very high [\ion{N}{ii}]/H$\textnormal{\alpha}$ intensity ratio for the optical nebula and also that the H$\textnormal{\alpha}$ filters used in the most of the previous observations also cover the [\ion{N}{ii}]$\textnormal{\lambda}\textnormal{\lambda}$6548,6583 lines within their bandpass response ranges (see Figure~\ref{fig:bandpasses} for details).

\subsection{The nature of the optical nebula}
The optical nebula observed at the location of the CCO in Vela Jr. SNR has sparked debate regarding its nature since its discovery over 20 years ago. Extensive radio and X-ray studies have not detected any PWN associated with the CCO \citep[e.g.][]{Pavlov2001,Reynoso2006,Maxted2018}. PWNe are expected typically to be powered by the rotational energy loss of an underlying pulsar or neutron star and, when young and still inside their SNR, they are confined by the pressure of the surrounding SNR material. They are also typically discovered in radio and X-rays. The lack of a PWN around the Vela Jr. CCO is consistent with the lack of PWNe in the CCO class. 

Our MUSE observations have revealed that the morphology of the optical nebula, particularly in H$\textnormal{\alpha}$, is reminiscent of a bow shock nebula, which is typically observed when pulsars move supersonically in their surrounding medium creating a nebula that is confined by the ram pressure in the ISM \citep[see e.g.][]{Gaensler2006}. The bow-shock-like structure is at approximately 3\arcsec away from the neutron star, which at a distance of 1 kpc, corresponds to about 0.015 pc. This distance is comparable to termination shocks seen in other PWNe. 
Such nebulae are characterized by being non-radiative and dominated by H$\textnormal{\alpha}$ Balmer line, with no detectable lines from any other element in the optical band. Although \citet{Mignani2019} dismissed the bow shock nebula scenario based on HST observations, our analysis using MUSE data suggests that this scenario cannot be excluded. Bow shock nebulae provide a valuable tool to probe the shock physics and can be used to constrain the underlying pulsar (or neutron star)'s spin-down energy $\dot{E}$ calculation.
From X-rays observations of the Vela Jr. SNR, an upper limit on the ambient density for thermally emitting plasma with $kT$$\sim$0.5 keV is established as $n_0 < 2.9 \times 10^{-2}$ cm$^{-3}$\citep{Slane2001}. With the electron density set as 4.8$n_0$, we estimate the thermal pressure $2n_e kT$ to be $\sim$$2.096 \times 10^{-10}$ erg cm$^{-3}$. Considering the standoff distance from the CCO to the H$\textnormal{\alpha}$ bow shock nebula determined through MUSE data to be approximately 0.015 pc, and assuming a spherical shape, as well as that the ram pressure of the wind termination shock is balanced by the surrounding thermal pressure, the estimated upper limit for any hidden pulsar's $\dot{E}$ is $1.69 \times 10^{35}$ erg s$^{-1}$.

\citet{Mignani2019} set an upper limit of 300 mas yr$^{-1}$ for the proper motion of the Vela Jr. CCO, based on a comparison of Chandra observations taken 9 years apart. However, a recent study of SRG/eROSITA and XMM-Newton data of the remnant has obtained a new coordinate for the geometric centre of Vela Jr., showing that the current position of the CCO is consistent with zero proper motion \citep{Camilloni2023}. This suggests that the CCO has likely remained in its birthplace without significant displacement. 

The morphology of the optical nebula observed in the other emission lines, as well as the overabundance of nitrogen, indicates a WR circumstellar nebula, which consist of material ejected by a massive star during its mass loss phase, that is currently being ionized by the Vela Jr. CCO. The circumstellar nebulae commonly found in young SNRs are typically characterized by slow expansion velocities and elevated [\ion{N}{ii}]/H$\textnormal{\alpha}$ ratios \citep[e.g.][]{Hartigan1987,Chu2001}. It has been proposed that Vela Jr. CCO is a product of the collapse of a massive star \citep{Slane2001,Allen2015}, and the recent eROSITA observation of the SNR supports a massive progenitor \citep{Camilloni2023}, although no specific mass or mass range had been inferred in these studies. Our test model, with a 60\,M$_{\sun}$ WN star as the progenitor using MAPPINGS~V, has provided circumstantial evidence that an object with a blackbody temperature similar to Vela Jr. CCO can be sufficient to create an optical nebula that is N-rich, H-weak, O-poor, and still shows faint emission from ionized heavy elements, much like what we observe.
Hence, it is reasonable to suggest that the progenitor is a WN star, that generated a nitrogen-rich circumstellar nebula during its mass loss. This nitrogen-rich nebula would have been subsequently shocked by the SN explosion and it is then being ionized by the CCO, leading to the nitrogen-abundant optical nebula we observe today. To validate our hypothesis, further modeling is needed, and we intend to address this in our forthcoming paper. 
If confirmed, this would be the first detection of a WR nebula around a CCO.

\section{Summary and Future Work} \label{sec:summary}
In this article, we have presented and discussed the results of a deep optical integral field spectroscopy observation, obtained with VLT's MUSE, of the optical nebula spatially coincident with the CCO in the Vela Jr. SNR. The sub-arcsecond spatial resolution of MUSE, combined with its AO capabilities, enables us to spatially-resolve, for the first time, the emission distribution of the long-debated optical nebula and characterize its emission line spectrum as being dominated by [\ion{N}{ii}] emission. The MUSE observation enabled the following key findings:
\begin{enumerate}
    \item The optical nebula exhibits a significant overabundance of [\ion{N}{ii}]$\textnormal{\lambda}\textnormal{\lambda}$6548,6583 emission, approximately 33 times the intensity of H$\textnormal{\alpha}$ emission.
    \item We identified a complex set of spectral lines associated to the optical nebula, including the nitrogen auroral line, sulfur lines, and the emission lines of the products of advanced stages of stellar evolution.
    \item The nebula displays two distinct morphologies: an arc-like shape that resembles a bow-shock nebula (observed in H$\textnormal{\alpha}$) and smooth blob structures (observed in [\ion{N}{ii}], [\ion{S}{ii}], [\ion{Ar}{iii}], [\ion{Fe}{ii}], and [\ion{S}{iii}]; and some with kidney-bean shape, observed on [\ion{N}{i}], and \ion{He}{i} emissions), while there is no nebulosity structure detected in the H$\textnormal{\textnormal{\beta}}$ and [\ion{O}{iii}] lines.
    \item We observed spatial structuring within the optical nebula, including a clear stratification of the ionized nitrogen levels.
\end{enumerate}

The MUSE clear detection of the distinct morphology of the optical nebula in H$\textnormal{\alpha}$, which was previously unobserved, has revived the bow shock nebula interpretation. The bow shock nebula scenario was dismissed based on the non-detection of the nebula in the HST WFPC2 F656N data \citep{Mignani2009} and the very minor contribution from H$\textnormal{\alpha}$ in the nebula's spectrum \citep{Mignani2019}. The overabundance of nitrogen in the optical nebula is consistent with a WR star progenitor and with the notion of the Vela Jr. SNR being associated with a massive progenitor. Our preliminary test model using \textsc{MAPPINGS}~V (i) has demonstrated the feasibility of a massive star in its WR stage producing an optical nebula akin to the one observed in Vela Jr. CCO and (ii) supports the suggestion that very massive progenitors not necessarily making black holes.

To gain a deeper understanding of the physical conditions within the nebula, a comparison of our observations with models is needed, particularly to simulate the heating and cooling processes that lead to the observed emissions. Photoionization models such as \textsc{MAPPINGS} 
and \textsc{CLOUDY} \citep{Ferland2013} solve the ionization and energy equilibrium equations, calculate the radiation transfer, and provide estimates of the physical quantities and the line intensities expected from a specific gaseous nebula. Through modeling, we also seek to comprehend the origins of the concentrated emissions of sulfur, argon, and iron within the nebula, and how they fit to the optical nebula interpretation. 
A more detailed modeling of the optical nebula's spectra using photoionization codes is deferred to a future publication.

\section*{Acknowledgements}

This work made use of NASA's Astrophysics Data System and various \textsc{Python} packages: \textsc{Numpy} \citep{harris2020array},
            \textsc{Scipy} \citep{2020SciPy-NMeth},
            \textsc{Astropy} \citep{astropy:2013, astropy:2018, astropy:2022},
            \textsc{brutifus} \citep{brutifus} (a Python module to process datacubes
                from integral field spectrographs, that relies on \textsc{statsmodel} \citep{seabold2010}, \textsc{matplotlib}, \textsc{astropy}, and \textsc{photutils}, an affiliated package of \textsc{astropy} for photometry), 
            \textsc{Matplotlib} \citep{Hunter:2007} and specifically \textsc{cubehelix} colormap \citep{cubehelix}, and
            \textsc{APLpy} \citep{aplpy}.
JS acknowledges funding from the University of Manitoba Graduate Fellowhips (UMGF) and the Ernst and Ingrid Bock Graduate Award. SSH acknowledges support from the Natural Sciences and Engineering Research Council of Canada (NSERC) through the Discovery Grants and the Canada Research Chairs programs, and by the Canadian Space Agency. CJL acknowledges the support of the NSTC grants 110-2112-M-001-020 and 111-2112-M-001-063. AJR was supported by the Australian Research Council through award number FT170100243. 

\section*{Data Availability}

The Multi Unit Spectroscopic Explorer (MUSE) data can be downloaded from the ESO Science Archive Facility at http://archive.eso.org/cms.html, under programme ID 0104.D-0092(B), P.I.: F.P.A. Vogt.



\bibliographystyle{mnras}




\appendix


\bsp	
\label{lastpage}
\end{document}